\documentstyle[11pt]{article}
\renewcommand{\theequation}{\thesection.\arabic{equation}}
\def\1{\mbox{1\hspace{-.8ex}1}}

\def\g{\gamma}

\def\d{\delta}
\def\e{\epsilon}

\def\ve{\varepsilon}
\def\f{\varphi}

\def\k{\kappa}

\def\lb{\label}

\def\m{\mu}
\def\n{\nu}

\def\r{\rho}

\def\ub{\mbox{\boldmath$\ub$}}

\def\ub{\mbox{\boldmath$u$}}
\def\p{\partial}

\def\pib{\mbox{\boldmath$\pi$}}

\def\z{\zeta}
\def\bi{\bibitem}

\def\B{\begin{equation}}
\def\E{\end{equation}}

\begin{document}

\thispagestyle{empty}

\begin{flushright}   hep-th/9904003 \\
                    LPTENS 99/09  \\
		    ULB-TH/99-07 \end{flushright}

\vspace*{0.5cm}

\begin{center}{\LARGE {Noether superpotentials in supergravities}}

\vskip1cm

M. Henneaux$^{a,b}$, B. Julia$^c$  and S. Silva$^c$

\vskip0.2cm

$^a$Physique Th{\'e}orique et Math{\'e}matique, Universit{\'e} Libre de
Bruxelles, \\
Campus Plaine C.P. 231, B--1050 Bruxelles, Belgium\\

\vskip0.1cm

$^b$Centro de Estudios Cient{\'\i}ficos de Santiago,\\
Casilla 16443, Santiago 9, Chile \\

\vskip0.1cm

$^c$Laboratoire de Physique Th{\'e}orique CNRS-ENS\\ 
24 rue Lhomond, F-75231 Paris Cedex 05, France\footnote{Unit\' e
propre du CNRS, associ\' ee \` a l'Ecole Normale Sup\' erieure et \` a
l'Universit\' e de Paris-Sud. This work has been partly supported by
the EU TMR contract ERBFMRXCT96-0012.}
\vskip1.0cm

\begin{minipage}{12cm}\footnotesize

{\bf ABSTRACT}
\bigskip

Straightforward
application of the standard Noether method in supergravity theories yields
an incorrect superpotential for local supersymmetry transformations,
which gives only half of the correct supercharge.  We show how to derive
the correct superpotential through Lagrangian methods, by applying
a criterion proposed recently by one of us.  We verify
the equivalence with the Hamiltonian formalism.  It is also
indicated why the first-order and second-order
formalisms lead to the same superpotential.
We rederive in particular the central extension by the magnetic charge
of the ${\cal N}_4 =2$ algebra of SUGRA asymptotic charges.


\bigskip
\end{minipage}
\end{center}
\newpage

\section{Introduction}
\setcounter{equation}{0}

The problem of defining meaningful conserved charges in gauge theories
is notoriously subtle. It has been addressed in the literature
along various lines.
One approach relies on the use of Noether identities
and conserved currents \cite{No,BH2,Fl,Bg,Ju,BCJ,JS}.
Another is based on Hamiltonian methods and asymptotic symmetries
\cite{ADM,RT,Asht,AD}.  The first approach is probably the
most familiar and emphasizes locality.  It has been recognized,
however, that it suffers from ambiguities, which, if
improperly resolved, may lead to incorrect results.
It was recalled in \cite{JS} that in modern language, Noether already
showed that on-shell the conserved charge associated to one-parameter
subgroup of a gauge group is topological and hence lives at any infinity
(ignoring singularities). More than $40$ years later \cite{ADM,RT}, it was
understood that charges can indeed be defined at infinity for a good
choice of boundary conditions, and in one to one correspondence with
their (asymptotic) symmetries. The later may be finite or infinite in
number.

The purpose of this paper is to derive the correct superpotential
for the supercharges in supergravity theories.  
We show that a ``natural''  application of
the Noether identities yields an incorrect supercharge.  We then
derive a correct superpotential by adopting the
criterion proposed in \cite{Si} and verify equivalence with
the Hamiltonian approach.   
Finally, we explain in a first appendix why the first-order and
second-order formulations yield the same superpotentials for
local supersymmetries.  The second appendix analyses in
some detail the case of ${\cal N}_4 =2$ supergravity and provides
explicit boundary conditions for the fields that enable one
to meaningfully compute the charges and their algebra that contain a
central charge as in the rigid supersymmetry case.

\section{Noether superpotential}
\setcounter{equation}{0}

As shown by E. Noether,
any  continuous one-parameter invariance of the action leads to a conservation law
$\partial_\mu j^\mu \approx 0$, where $\approx$ means ``{\it equal to}  
when the equations of motion
hold''.  The conserved Noether current is defined through
\B j^\mu = S^\mu - \frac{\partial L}{\partial \partial_\mu \phi}
\delta \phi \label{NC} \E
where $\delta \phi$ is the infinitesimal variation under the (local or
global) symmetry
and 
$\delta L = \partial_\mu S^\mu$. The fields $\phi$ may carry an index
over which one sums in (\ref{NC}), but this will not be explicitly
indicated.  In the class of Lagrangians
having the same (bulk) Euler-Lagrange derivative ($L \rightarrow
L + \partial_\mu k^\mu$), one must adjust the
surface term in such a way that the action $\int d^Dx L = 0$
is truly stationary on-shell. 
This surface term is fixed by a choice of boundary conditions.

However, even for a given $L$, there is some ambiguity in the
choice of $S^\mu$ since the addition to $S^\mu$ of the
divergence $\partial_\nu k^{\mu \nu}$
of an antisymmetric tensor $k^{\mu \nu} = - k^{\nu \mu}$ does not modify
$\partial_\mu S^\mu$.  Expressed in terms of the currents,
this (topological) ambiguity reads
\B j^\mu \rightarrow j^\mu + \partial_\nu k^{\mu \nu} \label{redef}\E
and is particularly relevant in
the case of gauge symmetries.

Indeed, in this case, conserved Noether currents
$j^\mu$ derive from superpotentials,
\B  j^\mu \approx \partial_\nu U^{ \mu\n} , \;
U^{\m \nu} = - U^{\nu \mu} \E
This has been proved in many references (see \cite{No,BH2,Fl,Bg,Ju,BCJ,JS} and
also \cite{BBH} for a cohomological interpretation).  The fact that
$j^\mu$ derives from a superpotential implies that one can set
it equal to zero by means of the redefinitions (\ref{redef}).
In other words, ``everything is in the superpotential", which indicates
how crucial it is to resolve correctly the above-mentioned
ambiguities.  A wrong choice would lead to an incorrect
integrated charge,
\B Q = \int_V d^{D-1}x \, j^0 \approx \int_{\partial V} d^{D-2} S_i\, U^{0i} 
\E
which would e.g. not generate the appropriate transformations
through the Poisson bracket.

A ``natural'' choice for the superpotential
may seem to be
\B U^{\mu \nu} = -\frac{1}{2} (M^{\mu \nu}_A - M^{\nu \mu}_A) \xi^A 
\label{wrongsup} \E
with
\B M^{\mu \nu}_A = \frac{\partial L}{\partial \partial_\mu \phi}\Delta_A^\nu 
\label{w2} \E
One finds 
\B  j^\mu = \partial_\nu U^{\mu\nu} + \frac{\delta L}{\delta \phi}
\Delta_A^\mu \xi^A \label{current} \E
where the $\Delta_A^\mu$'s are the coefficients of the derivatives of
the gauge parameters in the variations of the fields,
\B \delta_\xi \phi = \xi^A \Delta_A + \partial_\mu \xi^A \Delta_A^\mu \lb{deltafi}\E
(we assume for simplicity that only the first-order derivatives appear).
The choice (\ref{wrongsup}) corresponds to taking the coefficient of $\partial_\nu \xi^A$
in $S^\mu= \xi^A \Sigma_{A}^{\mu }+\partial_{\nu} \xi^A \Sigma_{A}^{\mu \nu }$ to be symmetric in $\mu$, $\nu$ and may be regarded
as being ``natural" on this ground.  

However, this choice is not
always correct and does in fact give an incorrect supercharge in supergravity
for instance.

To see this, note that only the fields that transform
into derivatives of the gauge parameter 
contribute to the superpotential, and only the piece of the action containing
derivatives of these fields is relevant.  
In all supergravity theories, the relevant part of the supersymmetry
transformations is thus
\B \delta_\epsilon \psi^A_\sigma = \partial_\sigma  \epsilon^A +
\hbox{ ``more"} \E 
and the relevant piece in the action
is the kinetic term for the gravitini\footnote{The conventions are the
following: 
$\eta^{ab}=\{ -,+,\dots,+\}$, $\e_{01\ldots (D-1)}=1$, $\g^a$ are $D$ real Majorana 
matrices and $\g^{a_1\ldots a_i}:=\g^{[a_1}\ldots\g^{a_i]}$ .
The gravitini are described by Majorana spinors.},
\B \frac{i}{2} \bar{\psi}^A_\lambda \g^{\lambda \mu \nu}
\partial_\mu 
\psi^A_\nu. \E
One finds by application of  formulas (\ref{wrongsup}-\ref{w2})
\B U^{ \mu\nu}_{\bar{\e}} =- 
\frac{i}{2} \bar{\e}^A  \g^{ \mu \nu \lambda} \psi^A_\lambda. \label{incorrect}\E
As discussed in the first appendix, one obtains (\ref{incorrect}) 
by working either in first or
second order formalism.  

Although simple, the formula (\ref{incorrect}) is incorrect.  It gives only
half of the supercharge as can be seen by comparing with the Hamiltonian 
formalism \cite{T}
or by computing the variation of the supercharge
under a supersymmetry transformation, where one finds only half of the
$4$-momentum $P^\mu$ instead of $P^\mu$ itself.  
The correct supercharge is \cite{T}
\B Q_{\bar{\e}} =- i \int_{S_\infty} d^{D-2}S_i \; 
\bar{\e}^A \g^{0ik} \psi^A_ k 
\label{supercharges}\E
while the integral of $j^0$ given by 
(\ref{current}), (\ref{incorrect})
is clearly only half of this expression.
That (\ref{incorrect}) is 
incorrect is perhaps not surprising since it is well appreciated
that there exist in general relativity a plethora of superpotentials,
many of which yield incorrect energy, momentum, or angular momentum
\cite{Bg,Ko}.  For a recent and informative discussion,
see \cite{KBL}.
We have just 
pointed out the supersymmetric extension
of this problem.  What is needed is a criterion that selects
among the many candidate superpotentials the correct one.  Such
a criterion has been proposed in \cite{Si} and tested with
success in many models.  We apply below this criterion and show
that it yields the correct supercharge.

\section{Construction of the correct surface integrals at infinity}
\setcounter{equation}{0}

The approach proposed in \cite{Si} is a ``superpotential-based 
generalization" of the Hamiltonian approach of \cite{RT}.
It may not always be equivalent to it, but in the
case of theories like supergravity where one can write the
Lagrangian in terms of forms and exterior products (in the sense
specified after equation (\ref{defvm})), it does 
yield the same supercharge.

The starting point of \cite{Si} is
the relationship between the superpotential and the conserved
current associated with a given one parameter group of gauge 
 transformations of the fields
through the Noether identities.
For any choice of surface terms, this relationship reads \cite{JS}
\B  j_U^\mu = \partial_\nu U^{ \mu\nu} + \frac{\delta L}{\delta \phi}
\Delta_A^\mu \xi^A \label{currentbis} \E
with the term proportional to the equations of motion being
independent of the choice.
The idea is then to find a criterion which, given the term
proportional to the equations of motion, completes it in a definite
way in equation (\ref{currentbis})\footnote{The current in
(\ref{currentbis}) is conserved on-shell due to the antisymmetry of $U^{\m\n}$
and the so-called Noether identities, $\p_\m \left(\frac{\delta L}{\delta \phi}
\Delta_A^\mu \xi^A \right) = \frac{\delta L}{\delta \phi} \d_{\xi} \f$.}.

In the case of supergravity, 
the supersymmetry current identity is
\begin{eqnarray}
j_{\bar{\epsilon}}^\mu &=& \partial_\nu U^{ \mu\nu}_{\bar{\epsilon}}
+ \bar{\e}^A \frac{\delta L}{\delta \bar{\psi}_\mu^A} \nonumber \\
&=& \partial_\nu U^{ \mu\nu}_{\bar{\epsilon}}
+ i \bar{\e}^A  \g^{\mu \rho \sigma} (\partial_\rho
\psi^A_\sigma + \Lambda_{\rho \sigma}^A)
\lb{jm} \end{eqnarray}
where $\Lambda_{\rho \sigma}^A $ 
denotes terms containing undifferentiated gravitini
fields.

If one varies this equation with respect to the gravitini
fields\footnote{In computing the superpotential associated with
supersymetries, one may assume $\d$(other fields)$=0$,
since the terms proportional to $\d$(other fields) die off
faster at infinity, where the superpotential is actually defined.
For instance, in four dimensions, this condition 
is verified
in the asymptotically flat case, for which we adopt the
boundary conditions of appendix B, or in the asymptotically
anti-de Sitter case, for which we take the precise boundary
conditions of \cite{HTads}.  Thus the expression (\ref{superpotential})
for the superpotential associated with supersymmetries
is correct in both cases.},
one gets, upon integration by parts,
\B \delta j_{\bar{\epsilon}}^\mu =
\partial_\nu \delta U^{ \mu\nu}_{\bar{\epsilon}}
+ \partial_\nu V^{ \mu\nu} 
- i \partial_\nu \bar{\e}^A  \g^{\mu \nu \sigma} \delta \psi^A_\sigma 
+ i \bar{\e}^A \g^{\mu\rho \sigma} \delta \Lambda_{\rho \sigma}^A
\label{divergence} \E
with
\B V^{\mu\nu} (\epsilon, \phi, \partial \phi)
= i \bar{\e}^A  \g^{\mu \nu \sigma} \delta \psi^A_\sigma
\lb{defvm}\label{defV}   \E
 Note that here $V^{\mu \nu}$ is antisymmetric in $\mu$ and
$\nu$. In fact, $V^{\mu\nu}$ is defined by (\ref{divergence}) up to
a total divergence, $\tilde{V}^{\mu \nu}=V^{\mu \nu}+\p_\r
X^{\m[\n\r]}$, which can in the present case be adjusted so
that $V^{\mu\nu}$ is antisymmetric in
$\m$ and $\n$. Quite generally, the antisymmetry is guaranteed 
if the theory can be written in a first order formulation (see
appendix of \cite{Si}).

What is proposed in \cite{Si} is to take $U^{ \mu\nu}$
in such a way that the divergence terms cancel
in (\ref{divergence}), i.e., such that
\B \delta U^{ \mu\nu}_{\bar{\epsilon}}
+  V^{ \mu\nu} = 0 
\label{criterion} \E
The integration in field space of this equation is straightforward
and leads to
\B U^{\mu\nu}_{\bar{\epsilon}} = -
i \bar{\e}^A  \g^{\mu \nu \sigma}\psi^A_\sigma \label{superpotential}\E
without the factor one-half (recall that $\delta$(other fields) $=0$ -
as stated in \cite{Si}, (\ref{criterion}) must be imposed only at infinity). 
As in the Hamiltonian formalism of
\cite{RT}, equation (\ref{criterion}) defines the charge up to a constant
which can be adjusted so that the superpotential vanishes for 
the vacuum. This superpotential
is  correct
since, contrary to (\ref{incorrect}), it yields the correct supercharges
(\ref{supercharges}). The last two terms of (\ref{jm}) are not relevant 
for charge evaluation. Actually the bulk charge has not 
yet been defined for singular solutions even
in the presence of horizons.

That the supercharge that follows from (\ref{criterion}) is
the same as the one obtainable by Hamiltonian methods is
easy to understand.
Indeed, one may identify $\partial_i V^{0i}$ as the surface term 
that one picks up at {\em the spatial boundary} when
breaking the variation of the spatial integral of the field-equation
term in (\ref{currentbis})
\B \delta \int_{x^0=C^{t}} d^{D-1} x \,\frac{\delta L}{\delta \phi}
\Delta_A^0 \xi^A := \delta \int_{x^0= C^{t}}
d^{D-1} x \, \bar{\epsilon}^A \frac{\delta L}
{\delta \bar{\psi}_0^A} \E
into a bulk term and a surface term,
\B \delta \int_{x^0= C^{t}} d^{D-1} x \,\bar{\epsilon}^A \frac{\delta L}
{\delta \bar{\psi}_0^A} = \hbox{ ``bulk" } + \int_{S_\infty}
d^{D-2} S_i \;V^{0i}. \label{varbulk} \E
where ``bulk'' contains only undifferentiated variations of the canonical fields.
This identification manifestly holds in supergravity. 
The only field equation that appears in (\ref{varbulk})
is the equation associated with the field that transforms
into the time derivative of the gauge parameter, 
i.e. $\bar{\psi}^A_0$.

Now, it is well-known that the temporal component $\bar{\psi}^A_0$
of the gravitino field is the Lagrange multiplier for the
supersymmety constraint-generator, namely $\frac{\delta L}
{\delta \bar{\psi}_0^A}$.  Therefore, the term being
varied in (\ref{varbulk}) is of the form ``Lagrange
multipliers" times ``Hamiltonian constraints".
This property, verified here for supergravity,
is actually generic since the
Lagrange multipliers transform always into the time
derivatives of the gauge parameters (see e.g. \cite{HT}).
Thus, 
the zeroth component of the criterion (\ref{criterion})
\B \delta U^{0i}_{\bar{\epsilon}} + V^{0i} = 0 \E
precisely guarantees that the charge
\B Q_{\bar{\e}} = \int_{x^0= C^{t}}
d^{D-1}x \;  \bar{\epsilon}^A \frac{\delta L}
{\delta \bar{\psi}_0^A} + \int_{S_\infty} d^{D-2}S_i \,U_{\bar{\e}}^{0i} \E
has well-defined functional derivatives
in the sense of \cite{RT}, i.e., has a variation that
contains only a bulk part.  Since the asymptotic equality
$\delta U^{0i}_{\bar{\epsilon}} + V^{0i} = 0$, together with 
a covariance argument at infinity, implies
$\delta U^{\mu\nu}_{\bar{\epsilon}}
+  V^{ \mu\nu} = 0$ (asymptotically), one
may conclude that in the case of supergravity, the superpotential method
supplemented by the criterion of \cite{Si} and the Hamiltonian
method are equivalent.

\section{Conclusions}
\setcounter{equation}{0}

In this paper, we have shown how superpotential methods 
apply to local supersymmetry transformations in supergravity.
We have rederived the correct supercharges.
Once the supercharges are known, one may compute their
algebra  \cite{T,GH}.  
The case of ${\cal{N}}_4=2$ supergravity
is particularly interesting because the algebra of the
supercharges contains central charges.  One of these,
the magnetic central charge, arises
in exactly the same way as the conformal central charge
in $2+1$ anti-de Sitter gravity \cite{BH}: the central charge is not
seen in the algebra of the asymptotic symmetries, but does appear in the
algebra of their canonical generators.
The calculation is direct.  Because it lies
somewhat outside the main line of this paper,
it is discussed in the second
appendix where precise boundary conditions that include
magnetic sources are
displayed.

\section*{Acknowledgements}
\bigskip

We thank Eug{\`e}ne Cremmer and Thomas Materna
for useful conversations.
MH is grateful to the ``Laboratoire de Physique
Th{\'e}orique de l'Ecole Normale Sup{\'e}rieure" for
warm hospitality extended to him in May of 1998
while this work began.
The work of MH has been partly supported by the ``Actions de
Recherche Concert{\'e}es" of the ``Direction de la Recherche
Scientifique - Communaut{\'e} Fran{\c c}aise de Belgique", by 
IISN - Belgium (convention 4.4505.86) and by 
Proyectos FONDECYT 1970151 and 7960001 (Chile).

This work has been partly supported by
the EU TMR contract ERBFMRXCT96-0012.
\section*{Appendix A: Superpotential in first-order and second-order formalisms}
\renewcommand{\theequation}{A.\arabic{equation}}
\setcounter{equation}{0}

The superpotentials for pure gravity are better understood in a pure
$1^{st}$ order formalism, i.e. where the connection $\omega_{\mu b}^a$
is varied independently \cite{JS}. However, supergravities are usually
given in a $2^{nd}$-order formulation of the supersymmetry
transformations laws. That is the (spin)-connection is not treated as
an independent field, and then, its supersymmetry transformation law
is not required. The point
of this appendix is to check in the special case of ${\cal{N}}_4=2$
supergravity that the above computation
of the superpotential 
does not depend on which formalism ($1^{st}$ or
$2^{nd}$ order) we are using. This provides a consistency check for
\cite{Si}. The same consistency holds for pure gravity 
\cite{JS2}.

First of all, the starting point for computing superpotentials is equation
(\ref{currentbis}). In the special case of supergravity, the result
was given in (\ref{jm}). 
Now, the only difference between a $1^{st}$
or $2^{nd}$ order computation would come from a term proportional to
the equation of motion of the connection when computing $V^{\m\n}$
using the second term of the rhs of 
equation (\ref{currentbis}). However, if the $1^{st}$
order formulation of supergravity is such that the supersymmetry
variation of the connection does not contain any term proportional to
$\p_\mu \xi^A$, no additional contribution is expected (recall
the definitions (\ref{deltafi}) and (\ref{currentbis})).  

In a parallel publication \cite{HJS2}, 
we will present a general scheme to derive a
pure $1^{st}$ order formulation from a  $1.5$ one. That
simply requires to compute a supersymmetry transformation law for the
connection. Then, the simplest way to show that the superpotential
computation for $1^{st}$ order formulation is equivalent to the
$2^{nd}$ order one is to show that this supersymmetry transformation law does
not depend on the derivative of the gauge parameter. 
This holds for the ${\cal{N}}_4=1$ supergravity, where the supersymmetry
transformation law for the connection is \cite{DZ}:

\B \d_1 \omega_{\m}^{ab} = 2 i\k^2 \bar{\e} \g^5 \left(\g_\mu \tilde{\psi}^{ab} -
\frac{1}{2} e_\m^a \g_c \tilde{\psi}^{cb}
+ \frac{1}{2} e_\mu^{b} \g_c \tilde{\psi}^{ca} \right) \lb{varomf} \E
where $ \tilde{\psi}^{ab} :=\frac{1}{2} \e^{abcd} {\cal D}_c \psi_d$, with the 
covariant derivative acting on spinors defined as usual by 
${\cal D}_{\sigma }:=\partial_{\sigma } + \frac{\gamma _{ab}}{4}\omega
_{\sigma }^{ab}$.

In the ${\cal{N}}_4=2$ supergravity case, it is also possible to find a
$1^{st}$ order formulation, that is a supersymmetry transformation
law for the connection. The result is  \cite{HJS2}:

$$ \d_1 \omega_\m^{ab} = 2 i\k^2 \bar{\e}^A \g^5 
\left(\g_\mu \tilde{\hat{\psi}}^{ab}_A -\frac{1}{2} e_\m^a \g_c 
\tilde{\hat{\psi}}^{cb}_A 
+ \frac{1}{2} e_\m^b \g_c \tilde{\hat{\psi}}^{ca}_A \right) $$

\B + i \k^3 \bar{\e}_A \ve^A_{\ B} \left(\hat{F}^{ab}- 
\tilde{\hat{F}}^{ab} \g^5 \right) \psi^B_\m \lb{tranw}\E
Here  $\tilde{\hat{\psi}}_{ab}^A := \frac{1}{2} \ve_{abcd} 
\hat{{\cal D}}^c \psi^{dA}$. The $( {\cal N}_{4}={2})$-``hatted'' covariant
derivative is defined by $\hat{{\cal D}}_{\sigma } :={\cal D}_{\sigma } +\frac{\kappa }{2}\ve^A_{\ B}\left(\hat{F}_{\rho \sigma } \gamma ^{\rho }+\tilde{\hat {F}}_{\rho \sigma }\gamma^{\rho } \gamma ^{5}\right)$ together with 
(super-covariant) ``hatted'' field strength $\hat {F}_{\mu
\nu }:= F_{\mu\nu }+i\kappa\ve_{AB} \bar{\psi}^{A}_{\mu }\psi^{B}_{\nu
}$ and its Hodge-dual $\tilde{\hat{F}}_{\mu
\nu }:=\frac{1}{2}\ve_{\mu \nu \rho \sigma } \hat{F}^{\rho \sigma}$.

Again, there is no term in $\p_\m \bar{\e}^A$ in the above
transformation law.
Another example is the eleven dimensional supergravity,
which will be presented also in \cite{HJS2}.

\section*{Appendix B: Boundary conditions, asymptotic symmetry algebra
and algebra of charges for ${\cal N}_4 = 2$ supergravity}
\renewcommand{\theequation}{B.\arabic{equation}}
\setcounter{equation}{0}

To compute the algebra of the charges of ${\cal N}_4 = 2$ supergravity,
one first needs to give boundary conditions for the fields
at spatial infinity that ensure that the surface integrals yielding the
corresponding charges be all finite.  The boundary conditions should also
be invariant under the asymptotic symmetries, i.e., under the
asymptotic rigid $N=2$ SUSY algebra.  In the second order formalism,
the independent fields of ${\cal N}_4 = 2$ supergravity are the tetrads
$e^a_\mu$, the abelian connection $A_\mu$ and the two gravitini
$\psi_\mu^A$.  The boundary conditions for the tetrads and the
gravitini are given in \cite{RT,T} (the tetrads approach their
Minkowskian values up to terms falling off like $1/r$ with definite
parity conditions, while the gravitini fall off
like $1/r^2$).  For this reason, we shall focus here only on the
abelian connection.  Our treatment includes
magnetic monopoles.

As in \cite{RT,T}, we shall express the boundary conditions
in terms of the canonical variables, which are $A_k$ and $\pi^k$
(electric field).  
To enforce the Lorentz invariance
properties of the boundary conditions, it is convenient to use
the ``improved" form of the transformation of the connection under
diffeomorphisms, which is
\B \delta_\xi A_\mu = \xi^\nu F_{\nu \mu} \label{potential} \E
This is in fact the transformation that arises in the
Hamiltonian formalism \cite{CT78}.
Indeed, the constraints associated with diffeomorphisms are, for
$N=2$ supergravity,
\begin{eqnarray} {\cal H} &=& {\cal H}^G + {\cal H}^{em} 
+ {\cal H}^{\frac{3}{2}} \approx 0 , \label{constraint1}\\
{\cal H}_k  &=& {\cal H}^G_k + {\cal H}^{em}_k
+ {\cal H}^{\frac{3}{2}}_k \approx 0 \label{constraint2} 
\end{eqnarray}
where ${\cal H}^G$ and ${\cal H}^G_k$ are the metric contributions, while
${\cal H}^{em}$ and ${\cal H}^{em}_k$ (respectively, ${\cal H}^{\frac{3}{2}}$
and ${\cal H}^{\frac{3}{2}}_k$) are the electromagnetic contributions
(respectively, the gravitini contributions).  We shall only need here
the electromagnetic terms (the gravitational terms do not contribute
to the transformations of the electromagnetic variables, while the
spin $3/2$ terms can be neglected asymptotically).  One has
\begin{eqnarray} {\cal H}^{em}&=& \frac{g_{ij}}{2 \sqrt{g}} (\pi^i \pi^j +
{\cal B}^i {\cal B}^j) \\
{\cal H}^{em}_k &=& F_{km} \pi^m
\end{eqnarray}
(electromagnetic energy density and Poynting vector).  Here, ${\cal B}^i$ is
the magnetic field and $g_{ij}$ the spatial metric induced on the
surfaces $x^0 = C^{t}$; $g$ is its determinant.  It is straightforward
to check that $\int d^3x (\xi {\cal H} + \xi^k {\cal H}_k)$, where $\xi$
is the displacement
normal to the hypersurface $x^0 = C^{t}$ and $\xi^k$ the tangential
displacement, generates (\ref{potential}) when acting on $A_k$
through the Poisson bracket.  The transformation law
(\ref{potential}) for $A_0$
follows then from the general variation of the
Lagrange multiplier under transformations generated by the
constraints (see e.g. \cite{HT} equation (3.26b) - note that the brackets
of the constraints (\ref{constraint1}) and (\ref{constraint2})
involve the Gauss' law constraint \cite{CT78}).

The transformation of the electric field will also be evaluated
through its Poisson bracket with the Hamiltonian constraints
and is
\B \delta_{\xi} \pi^k = D_m(\xi F^{km} \sqrt{g}) 
+ (\xi^m \pi^k),_m - \xi^k,_m \pi^m 
\label{electric} \E
where $D_m$ is the spatial covariant derivative.
The transformation (\ref{electric}) coincides on-shell with 
the standard transformation.

To motivate the boundary conditions, consider first the zero-monopole sector.
In Minkowski space, the electromagnetic potential for an 
electric charge at rest at the origin is given by
\B A_0 \sim \frac{e}{r}, \; \; A_k = 0 \E
while the electromagnetic field behaves as
\B F_{0i} \sim \frac{e n^i}{r^2}, \; \; F_{ij} = 0 \E
where $n^i$ is the unit radial vector.
By boosting this solution, one generates non-vanishing
$r^{-1}$-order terms for $A_k$;  these terms have
the interesting property of being even under the parity
${\bf n} \rightarrow -{\bf n}$.  Similarly, the leading-order
term in the electromagnetic field has the property of
being odd under the same parity transformation.  These
features remain valid if one superposes various charges.

This suggests taking as boundary
conditions at spatial infinity (at each given time)
\B A_k = \frac{a_k^{(1)}({\bf n})}{r} + \frac{a_k^{(2)}({\bf n})}{r^2}
+ o(r^{-2}) \label{bcp} \E
for the spatial components of the connexion and 
\B \pi^k = \frac{p^k_{(1)}({\bf n})}{r^2} + \frac{p^k_{(2)}({\bf n})}{r^3}
+  o(r^{-3}) \label{bcef} \E
for the electric field.  These conditions imply
\B F_{ik} = \frac{f_{ik}^{(1)}({\bf n})}{r^2} +
\frac{f_{ik}^{(2)}({\bf n})}{r^3} + o(r^{-3}) \label{bcmf}\E
for the magnetic field.
We  impose also that the first coefficients in $A_k$ and $\pi^k$
has definite parity properties:
\B a_k^{(1)}( - {\bf n}) = a_k^{(1)}({\bf n}), \; \;
p^k_{(1)}( -{\bf n}) = - p^k_{(1)}({\bf n}) . \E
It follows that
\B f_{ik}^{(1)}(-{\bf n}) = - f_{ik}^{(1)}({\bf n}) \E
One easily verifies that these conditions are invariant under
Lorentz transformations and futhermore guarantee the vanishing of the
boundary term of the variational principle, namely $\delta A_{0}
\pi^{i}+\delta A_{j} F^{ji}$, when integrated at infinity.  In fact, the combined asymptotic
conditions of \cite{RT,T} and those given here are invariant
under the full rigid $N=2$ SUSY algebra acting at infinity.

We shall verify for example the asymptotic Lorentz invariance
of the above boundary conditions.  The invariance of (\ref{bcp}) follows
from (\ref{potential}) and (\ref{bcef}), (\ref{bcmf})
and the fact that for boosts and rotations, the leading
orders of $\xi$ and $\xi^k$ are both parity-odd.  Similarly,
the invariance of (\ref{bcef}) follows from (\ref{electric})
and the same observation on the parity of the leading
orders of $\xi$ and $\xi^k$.  One should stress that the boundary
conditions adopted here can probably be somewhat relaxed.  However,
our goal is not to provide here the most flexible admissible
boundary behaviour, but to give a consistent and complete
set of boundary conditions that enforce asymptotically
the $N=2$ supersymmetry algebra.

The asymptotic symmetry algebra may actually  involve also a central
$U(1)$.  This issue is somewhat subtle because there is no charged
field in ${\cal{N}}_4=2$ (ungauged) supergravity.  Whether there is a non
trivial $U(1)$ at infinity depends on the topology of the spatial sections.
To see this, one notes that
the boundary conditions on the potential are invariant under
gauge transformations that behave asymptotically as
\B \delta_{\Lambda } A_\mu = \partial_\mu \Lambda \label{reducible} \E
with
\B \Lambda = \lambda_0 + \lambda({\bf n}) + o(r^0) \label{asympt} \E
where $ \lambda_0$ is a constant and 
$\lambda({\bf n})$ is parity-odd,
\B \lambda(- {\bf n}) = - \lambda({\bf n}). \E
However, the transformations associated with $\lambda({\bf n})$
are really irrelevant because their corresponding charges, given
by the flux of the electric field times $\lambda({\bf n})$
at infinity (up to non-written Gauss' law constraint terms)
\B Q[\lambda({\bf n})] = \int_{S_\infty} \lambda({\bf n})
 \pib \cdot {\bf dS} \label{irrel} \E
all vanish identically for the field configurations allowed here.
Thus, one can factor them out.
These transformations are somewhat the analogs of the supertranslations of the
BMS group, which are eliminated in the same way through the
parity conditions \cite{RT}.  In fact, the main motivation for
adopting similar parity conditions on the
electromagnetic potential is to effectively
remove the charges (\ref{irrel}), which do not seem to have a direct
physical interpretation (these charges are {\em not} associated with
higher multipole moments).

Thus, only the constant piece $\lambda_0$ in (\ref{asympt}) is
relevant.  
The charge
associated with gauge transformation that tend to a constant
at infinity reads (again up to unwritten Gauss' law terms)
\B Q[\lambda_0] = \lambda_0 \int_{S_\infty}
\pib \cdot {\bf dS} \label{electriccharge}. \E
There may be additional surface terms if there are other boundaries
(other asymptotic regions, or other surfaces on which boundary conditions
are required: horizons, surfaces surrounding singularities), 
but since we are interested here only in the asymptotic
symmetry generators in a single asymptotic region (called
``infinity''), we shall consider gauge transformations that
are zero except in a vicinity of this asymptotic region.  These
gauge transformations are correctly generated by (\ref{electriccharge}),
without any other surface contribution.  Being zero at the other
boundaries (if any) they certainly preserve any set of 
specific boundary conditions
given there.

Now, whether (\ref{electriccharge}) identically vanishes or not
depends on the topology of the spatial sections.  If these
are homeomorphic to $R^3$, then (\ref{electriccharge}) is
zero by Gauss' law and there is no non trivial $U(1)$ asymptotic
generator.  However, if the $2$-sphere at infinity is not
contractible,
(\ref{electriccharge}) needs not vanish (``charge without charge").
Furthermore, as a generator, (\ref{electriccharge}) acts non trivially
on the gauge-invariant Mandelstam variables
\B M= \int_\gamma dx^k A_k \E
where $\gamma$ is any path joining the asymptotic region being considered
to any other asymptotic region.  The variable $M$ is invariant under
``proper" gauge transformations, i.e., under gauge transformations
that vanish on all boundaries.  In that sense, it is an observable.
However, it is not invariant under global transformations characterized by
a non-vanishing $\lambda_0$ provided it ends outside the component of
infinity under consideration.  
One gets $\delta_{\Lambda } M = \lambda_0$, which implies
that $M$ is conjugate to the electric charge measured at infinity,
\B [M, \int_{S_\infty}  \pib \cdot {\bf dS}] = 1 \E 
(again, we do not write explicitly the constraint terms that accompany
the surface integral at infinity).  There is thus a non-trivial $U(1)$
generator at infinity, which generates asymptotic symmetries.

Note that one would also find $U(1)$ symmetries on the
other boundaries.  The sum over all boundaries of the
$U(1)$ charges is zero by Gauss' law (the flux lines can only end
on the boundaries).  
The fact that the charges add up to zero
just reflects the fact that the generator associated with everywhere
constant gauge transformations vanishes, as it should (such
transformations have no action on the fields).

The inclusion of
magnetic charges is direct.  Instead of imposing
(\ref{bcp}) on the potential $A_k$, one requires that it
behaves asymptotically as
\B A_k = A_k^{mon} + A'_k \E
where $A_k^{mon}$ is the potential for the magnetic monopole(s),
living on a non-trivial $U(1)$ bundle over $S_2$ (and defined
over patches), while $A'_k$ is globally defined and subject
to the same boundary conditions (\ref{bcp}).  The boundary
conditions on the electric field are unchanged.  The
magnetic field has also the same asymptotic behaviour because the field
of a monopole obeys (\ref{bcmf}).  These more general boundary
conditions are still Lorentz invariant because the variation
of $A_k^{mon}$, which is globally defined, behaves as $A'_k$.
They are in fact also invariant under supersymmetry transformations and constant
gauge transformations.  
Furthermore,  the (graded) commutator
of two local supersymmetry transformations that tend asymptotically
to constants $\epsilon^A$ and $\zeta^A$
is  the sum of a diffeomorphism that becomes asymptotically 
a translation with parameter $-i \kappa^2 \bar{\epsilon}_A
\gamma^\mu \zeta^A$ and a gauge transformation that
behaves at infinity as a constant gauge transformation
with parameter $\Phi = -i \kappa \bar{\epsilon}_A \varepsilon^A_B \zeta^B$
(the asymptotic value of the gauge transformation is defined
up to a constant, but the ambiguity disappears
if one imposes that $\Phi$ vanishes when  $\epsilon$ or
$\zeta$ is equal to zero).  There is no sign of the magnetic charge
in the asymptotic algebra of the asymptotic symmetry transformations.

Let us turn now to the algebra of the charges defined, in
the Lagrangian context, through the variations of the
charges under the transformations generated by one another.
The  variation of the superpotential was
given in equation (\ref{criterion}), together with
the result (\ref{defV}) for supergravity theories. Now
in the special case where $\d$ is another supersymmetry gauge
transformation, that is $\d:=\d_\z$, the variation becomes: 
\begin{eqnarray}\label{algebre}
\d_\z U_{\bar{\e}}^{\mu \nu } &=& -i \bar{\e}_A\gamma ^{\mu \nu \sigma } \hat{\cal
 D}_{\sigma }\z^A\nonumber\\
&=& -i \bar{\e}_A\gamma ^{\mu \nu \sigma } \left( {\cal D}_{\sigma } +\frac{\kappa }{2}\ve^A_{\ B}\left(\hat
{F}_{\rho \sigma } \gamma ^{\rho }+\tilde{\hat {F}}_{\rho \sigma }\gamma
^{\rho } \gamma ^{5}\right)\right)\z^B \nonumber\\
&=& -i \bar{\e}_A\gamma ^{\mu \nu \sigma }{\cal D}_{\sigma }\z^A
-i\kappa \bar{\e}_A \ve^A_{\ B} \left(\hat {F}^{\mu \nu } + \gamma
^{5}\tilde{\hat {F}}^{\mu \nu } \right) \z^B
\end{eqnarray}

Where we used the $( {\cal N}_{4}={2})$-``hatted'' covariant derivative defined in appendix A, after equation (\ref{tranw}).

Let us now integrate (\ref{algebre} ) at spatial infinity:
\begin{enumerate}
\item The first term in the rhs of (\ref{algebre}) is nothing but the
Nester-Witten superpotential \cite{NW}, and so the covariant version of
ADM mass (and momentum). We will denote it by 
$ H_{\xi }$, where 
$\xi^{\mu }$ is an asymptotic displacement and is given by $\xi^{\mu
}=-i \kappa ^{2}\bar{\e}_{A}\gamma
^{\mu }\z^{A}$. The idea of relating the positivity of gravitational
mass and local supersymmetry can be found in \cite{DTG}.
The supergravity origin and positivity of the
Nester-Witten superpotential \cite{NW} was already established for the
${\cal{N}}_{4}=1$ supergravity in \cite{Hu}.

\item The second term gives electric and magnetic contributions
(``central'' charges) to the algebra of supersymmetry. Due to the
asymptotic behavior of the gravitino (falling off like $1/r^{2}$), the
hatted field strength can be replaced by the ordinary field strength,
the difference between both falling off like $1/r^{4}$. 

\end{enumerate}

Let us define: 
\[
Q :=- \int_{S_{\infty}} \left( i \kappa \bar{\e}_A \ve^A_{\ B}\z^B
\right) F^{0i}\hspace{.15in}  \mbox{and}\hspace{.15in} P :=-\int_{S_{\infty}}
\left(i \kappa\bar{\e}_A \ve^A_{\ B}\g^5\z^B \right) \tilde{F}^{0i}\nonumber
\]
Then we find that the
algebra of the SUSY charges (which generate the asymptotic 
supersymmetric invariance of ${\cal{N}}_{4}=2$
supergravity) has a magnetic central
charge besides the expected $U(1)$ generator.
These are familiar from rigid supersymmetry.
Explicitly,  we find the algebra
\B \d_\z Q_{\bar{\e}}:=[Q_{\bar{\e}},Q_{\z}] =
H_{\xi} + Q +  P \label{susyAlgebra}\E
in agreement with \cite{T,GH}.  The electric and magnetic charges appear
as central charges for the $N=2$ supersymmetry algebra.  There is,
however, an apparent difference between the two in the present
(electric) formulation of ${\cal N}_4 = 2$ supergravity.  While the Noetherian
electric charge is a non trivial generator 
which does act on some canonical
fields (when it does not
vanish identically), 
the magnetic charge commutes with all of them and not just with the
other generators of the asymptotic symmetry algebra, even
when it is non zero.  It is
(in this limited treatment) a fully central charge. 

It is customary to adjust the constants in the generators
of the asymptotic symmetries so that these vanish on some
specified background.  When specialized to that background,
(\ref{susyAlgebra}) becomes
\B \d_\z Q_{\bar{\e}} = P  \hbox{ on background}\E
($H_{\xi} =0$, $Q =0$).  The central charge is the variation
of the background generator of asymptotic supersymmetries.
The magnetic central charge can (and does) arise because
the electromagnetic field on a non-trivial $U(1)$-bundle
is non zero and breaks some of the supersymmetries
($\d_\z Q_{\bar{\e}} \not= 0$).

Note that in a dual (magnetic) formulation the role of electric and magnetic 
charges would be exchanged, a fully duality invariant treatment should restore 
the symmetry between the two.

The algebra of the asymptotic symmetry
transformations (once more defined on the electric canonical variables)
and the algebra of their generators are different.
The algebra (\ref{susyAlgebra}) gives another example of the general theorem of
\cite{BH1} which establishes that extra central charges are allowed in the 
asymptotic charge algebras. This has been already used in \cite{BH} and
generalises the classical ambiguity of the charge algebra relative to the 
symmetry algebra in Hamiltonian dynamics.

\end{document}